# Supporting Defect Causal Analysis in Practice with Cross-Company Data on Causes of Requirements Engineering Problems

Marcos Kalinowski, Pablo Curty, Aline Paes
Computing Institute
Fluminense Federal University (UFF)
Niterói, Brazil
{kalinowski, pablocurty, alinepaes}@ic.uff.br

Alexandre Ferreira
Information Technology Area
Brazilian National Development Bank (BNDES)
Rio de Janeiro, Brazil
alefe@bndes.gov.br

Rodrigo Spínola
University of Salvador (UNIFACS) and
Fraunhofer Project Center at UFBA
Salvador, Brazil
rodrigo.spinola@fpc.ufba.br

Daniel Méndez Fernández
Software & Systems Engineering
Technical University of Munich
Munich, Germany
daniel.mendez@tum.de

Michael Felderer
Institute of Computer Science
University of Innsbruck
Innsbruck, Austria
michael.felderer@uibk.ac.at

Stefan Wagner
Institute of Software Technology
University of Stuttgart
Stuttgart, Germany
stefan.wagner@informatik.uni-stuttgart.de

*Abstract*—**[Context]** Defect Causal Analysis (DCA) represents an efficient practice to improve software processes. While knowledge on cause-effect relations is helpful to support DCA, collecting cause-effect data may require significant effort and time. **[Goal]** We propose and evaluate a new DCA approach that uses cross-company data to support the practical application of DCA. **[Method]** We collected cross-company data on causes of requirements engineering problems from 74 Brazilian organizations and built a Bayesian network. Our DCA approach uses the diagnostic inference of the Bayesian network to support DCA sessions. We evaluated our approach by applying a model for technology transfer to industry and conducted three consecutive evaluations: (i) in academia, (ii) with industry representatives of the Fraunhofer Project Center at UFBA, and (iii) in an industrial case study at the Brazilian National Development Bank (BNDES). **[Results]** We received positive feedback in all three evaluations and the cross-company data was considered helpful for determining main causes. **[Conclusions]** Our results strengthen our confidence in that supporting DCA with cross-company data is promising and should be further investigated.

I. INTRODUCTION

Defect causal analysis (DCA) [1] encompasses the identification of causes of defects and ways to prevent them from recurring in the future. Effective DCA has helped to reduce defect rates by about 50% in organizations such as IBM [2], Computer Science Corporation [3], and InfoSys [4]. Reducing defect rates also reduces rework and the total project effort, improving productivity [5].

Given these benefits, to provide guidance on how to implement DCA in software organizations, we conducted a Systematic Literature Review (SLR) in 2006 and replicated in 2007, 2009, and 2010 [5]. The results allowed producing evidence-based guidelines on how to implement DCA and the identification of investigation opportunities. For instance, we found that "the DCA state of the art did not include any approach integrating learning mechanisms regarding cause-effect relations into DCA meetings".

This result encouraged us to initiate research to propose a DCA approach integrating cause-effect learning mechanisms into DCA meetings. An initial concept [6] suggests aggregating knowledge gathered in successive within-company causal analysis events to assemble a deeper understanding of the defects' cause-effect relations by using Bayesian networks. It is noteworthy that Bayesian networks have been used in other contexts in the software engineering domain such as defect prediction [7] and decision making [8]. In the DCA context, their use aims at creating and maintaining common causal models to support defect cause identification in each DCA event [6]. The diagnostic inference of those causal models can then be used to answer questions during DCA meetings, such as: "Within my organizational context, what are the causes that commonly lead to a specific defect type?".

Later, this initial concept was evolved and tailored into the DPPI (Defect Prevention-Based Process Improvement) approach [9]. The experience of applying DPPI retroactively to a real software project indicated its usage feasibility [9]. Moreover, according to the participants, DPPI's Bayesian diagnostic inference predicted the main defect causes efficiently, motivating further investigation. This lead to an experimental study and the preliminary results indicated benefits of using the diagnostic inferences during DCA sessions regarding effectiveness and effort [10].

The positive results encouraged us to institutionalize the DPPI process in a small-sized software organization (~25

employees) [11]. There, the approach was applied to analyze defects of requirements and designs, allowing further comprehension on its industry readiness and objectively measuring effort and obtained benefits. The average application effort was reasonably low (15.5 hours) when compared to the obtained benefits (reducing defect rates by 46 percent for requirements and 50 percent for designs). While the defect rate results were similar to those mentioned in literature (e.g., [2][3][4][5]), the contribution of the Bayesian diagnostic inference to the DCA sessions was hindered by the absence of a great amount of within-company data on the defects' cause-effect relations.

More recently, we started investigating cross-company cause-effect relations for critical Requirements Engineering (RE) problems based on data from the NaPiRE (Naming the Pain in Requirements Engineering) project [12][13]. The NaPiRE project comprises the design of a globally distributed family of surveys to lay an empirical foundation on RE practice and problems [14]. The NaPiRE data provided us the opportunity to build a cross-company cause-effect model for practically relevant RE problems.

In this paper, we hypothesize that such cross-company cause-effect model could be useful to support DCA sessions in specific industrial contexts and empirically investigate this issue. Using cross-company data to support DCA sessions has not been considered so far. Our investigation followed the model for technology transfer suggested by Gorschek *et al.* [16]. This model consists of conducting consecutive evaluations in different contexts: (i) in academia, (ii) with industry representatives (*static validation*), and (iii) in a real industrial context (*dynamic validation*). The academic evaluation was conducted in a graduate course on software engineering. The evaluation with industry representatives was conducted with employees of the Fraunhofer Project Center at UFBA. Finally, we used our approach in an industrial case study conducted at the Brazilian National Development Bank. Results of all three evaluations indicate that the approach was well accepted. Moreover, the cross-company data was unanimously considered helpful for determining main causes during DCA.

The remainder of this paper is structured as follows. Section II presents the background on DCA and on the NaPiRE project. Section III describes how the cross-company cause-effect model was built. Section IV introduces the cross-company data supported DCA approach. Section V presents our industrial evaluation strategy. Section VI and VII describe the initial evaluations in academia and with industry representatives. Section VIII reports the industrial case study. Section IX discusses threats to validity before concluding our paper in Section X.

II. BACKGROUND

A. Defect Causal Analysis (DCA)

Card [17] summarizes the DCA process in six steps: (i) to select a sample of the defects; (ii) to classify selected defects; (iii) to identify systematic errors; (iv) to determine the main causes; (v) to develop action proposals; and (vi) to document meeting results. In this context, a systematic error is an error that results in the same or similar defects being repeated on different occasions. Finding systematic errors indicates the existence of improvement opportunities. Besides listing these six steps, he highlights the importance of managing the implementation of the action proposals until their conclusion and communicating the implemented changes to the development team.

A representation of the traditional software defect prevention process [18], consistent with the DCA process described above, is shown in Fig. 1. DCA can be considered part of defect prevention, which also addresses implementing improvement actions to prevent the causes (*action team* activity) and communicating changes to the development team (*stage kickoff* activity). The depicted experience base indicates defect prevention as a means for communicating lessons learned among projects.

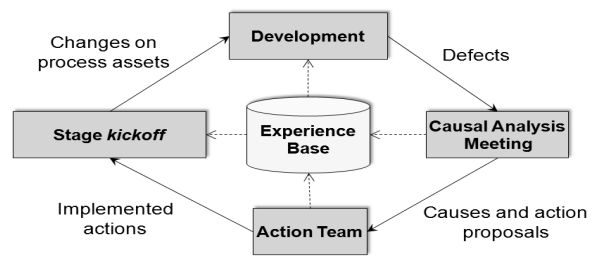

Figure 1. Defect Prevention Process. Adapted from [19].

Evidence-based guidelines for conducting DCA have been provided based on results of an SLR [5]. The guidelines provide advice on the techniques that can be used (e.g., Pareto charts for identifying categories to search for systematic errors or cause-effect diagrams for determining main causes for systematic errors), data to be collected, defect taxonomies, and cause categories. Also, general DCA assumptions and expectations (e.g., costs between 0.5 and 1.5 of the project budget and defect rate reductions of about 50%) are described.

B. The NaPiRE Project

The NaPiRE project comprises a globally distributed family of surveys to lay an empirical foundation on RE practice and problems [15]. The design of the survey instruments is aligned to a well-defined theory and has been extensively reviewed by over 40 researchers world-wide [14]. Further information on the project, including a sample of the questionnaire and the countries in which the survey is being replicated can be found online[1]. Concerning its results, four publications have specifically focused on causes of RE problems [12][13][15][19].

In [12], we presented the aggregated data on common causes for critical RE problems, based on answers from 74 Brazilian organizations. In [19], we focused on addressing the causes of the specific problem *incomplete/hidden requirements*, aiming at providing suggestions for its prevention based on data from Austria and Brazil. A comparison between RE practices, main problems and causes in Brazil and Germany can be found in [15]. Finally, in [11], we conducted a cause-effect investigation for critical RE problems based on data from the whole dataset [13].

---

[1] http://www.re-survey.org

The cross-company cause-effect model for RE problems built for this research is based on answers from the 74 Brazilian organizations (each one providing recurring causes for up to five RE problems they considered critical in their context). These 74 organizations include small, medium and large-sized organizations enrolled in both plan-driven and agile development. Further details on the Brazilian NaPiRE dataset can be found in [12].

Table I shows the five most critical RE problems, as ranked by the 74 organizations. Those were the only problems cited by more than 25% of the respondents. The table also shows how often each problem was ranked as the most critical. We observed that communication problems were often cited (problems #1 and #4), as well as incomplete and underspecified requirements (problems #2 and #3).

TABLE I.  MOST CRITICAL RE PROBLEMS

| # | RE Problems | Cited | Rank #1 |
|---|---|---|---|
| 1 | Communication flaws between the proj. team and the customer | 32 (43%) | 9 (12%) |
| 2 | Incomplete and/or hidden requirements | 31 (42%) | 12 (16%) |
| 3 | Underspecified requirements (too abstract) | 31 (42%) | 3 (4%) |
| 4 | Communication flaws within the proj. team | 26 (35%) | 5 (7%) |
| 5 | Insufficient support by customer | 21 (28%) | 5 (7%) |

### III. CROSS-COMPANY CAUSE-EFFECT MODEL FOR RE PROBLEMS

This section explains how the cross-company model on cause-effect relations was built. First, we prepared the NaPiRE data provided by the 74 Brazilian organizations and then built the cross-company cause-effect model as a Bayesian network based on that data using the Netica[2] tool.

We decided to concentrate on the five most cited problems shown in Table I. Those problems were cited (overall) 141 times. For each of these citations, the respondents indicated the main causes and effects in free text format. We analyzed and coded the textual data using the constant comparative method [20] and peer reviewed the codes. As a result, we created 49 cause and 41 effect codes.

Thereafter, we organized the information in a spreadsheet with 105 columns (5 for the problems, 49 for the cause codes, 5 for the cause categories, 41 for the effects, and 5 for effect categories) and 141 lines (one for each time a problem was cited by one of the subjects). Finally, for each line we marked the cited problem, its causes, cause categories, effects and effect categories. This was the data used for building the Bayesian network.

A Bayesian network consists of two related components: a qualitative one, which are the nodes representing random variables and the edges representing a direct relationship between two random variables (a direct acyclic graph - DAG); and a quantitative component, which are the conditional probabilistic distributions (CPDs) [21]. While an expert of the domain is able to specify the DAG, the same cannot be easily done with the CPDs.

---

[2] https://www.norsys.com/

We built the DAG in Netica by automatically adding random variables for each spreadsheet column and then manually connecting the variables with arrows indicating the causal relations between them. Thereafter, we used Netica to learn the CPDs from the spreadsheet data with the Expectation Maximization (EM) algorithm [22], which is indicated when the dataset includes empty values (as ours, given that some participants might not have indicated any causes) to approximate the CPDs to the real distribution.

The structure of the resulting Bayesian network DAG is shown in Fig. 2. The five cause categories grouping the causes were the ones suggested by the guidelines [5] (*input*, *method*, *organization*, *people*, and *tools*). For the effects, we defined categories that allowed us to consistently group the effects (*customer*, *design or implementation*, *product*, *project or organization*, and *verification or validation*).

Having built the Bayesian network, Netica allows conducting diagnostic and predictive inferences. In this paper, we focus on diagnostic inferences from problems to causes. The dynamics of this usage within our DCA approach will be detailed in the next section.

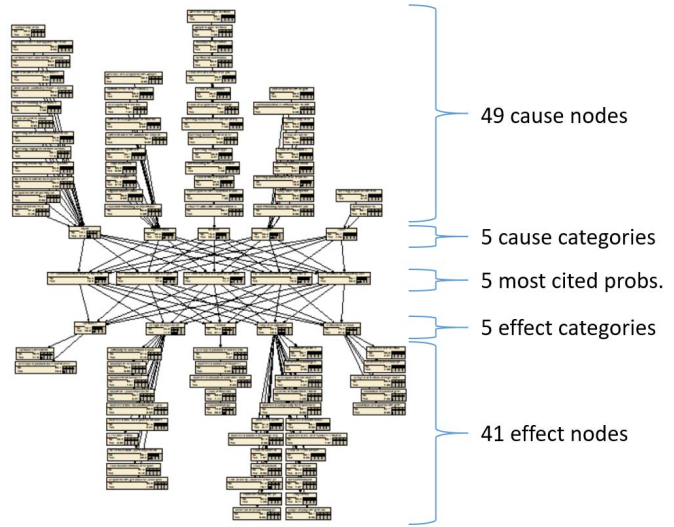

Figure 2. Structure of the Bayesian network DAG.

### IV. CROSS-COMPANY DATA SUPPORTED DCA

This section describes the proposed DCA approach, supported by the cross-company model on cause-effect relations of RE problems. In summary, the approach consists of performing the DCA steps detailed by Card [17], following the DCA guidelines [5] and supporting one of the steps, the determination of the main causes, with the cross-company model. Further details on how to perform each of the steps follow:

- *Select a sample of the defects*. The sample is represented by the defects found during the last inspection. We suggest using all the defects as the sample and filtering them by the main categories as part of the next step. However, in case of many defects a U-chart can be used to allow identifying data that deserves special attention [5].

- *Classify selected defects.* Following the guidelines, the nature of the defects [23] (*ambiguity, extraneous information, inconsistent information, incorrect fact,* and *omission*) is used as classification scheme. A Pareto chart is then applied to identify the most representative categories of defects, in which the identification of systematic errors takes place.
- *Identify systematic errors.* This step involves reading the defects of the most representative categories looking for similar defects being repeated in different occasions and the systematic errors leading to them.
- *Determine the main causes.* This step represents the core of DCA. The systematic errors are analyzed (ideally with representatives of the document authors, inspectors, and process group members [5]) seeking for causes that explain why they are happening. According to the guidelines, causes belong to one of the following categories: *input, method, organization, people,* and *tools*. This step is where we innovate by supporting the determination of the main causes with diagnostic inferences of a cross-company cause-effect model. These diagnostic inferences indicate causes that commonly lead to certain defect types and which could be considered as a starting point when brainstorming causes.
- *Develop action proposals.* Based on the causes, actions are proposed to help preventing the systematic error to reoccur in the future.
- *Document the meeting results.* This step assures that the results (e.g., causes and actions) are registered.

To illustrate the dynamics of using the network's diagnostic inference to support DCA, we use a simple example. Suppose that an organization has found several defects of type omission and that the systematic error is omitting details of business rules that are complex to understand. This systematic error regards the problem *incomplete/hidden requirements*. Therefore, before starting to discuss the causes from scratch, the DCA team could look at the Bayesian diagnostic inference.

This can be easily done with Netica by clicking on the selected problem. Given the size of the Bayesian network, we show excerpts of it that are relevant for the example. The diagnostic inference of this initial step is shown in Fig. 3. Based on the Bayesian probabilities it is possible to observe that the causes for this problem usually fall into categories *people* (70%), *input* (40%), or *method* (30%).

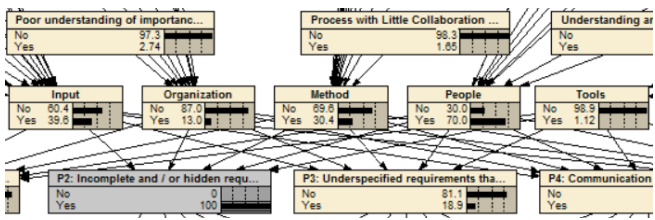

Figure 3. Diagnostic inference for causes after selecting problem *incomplete/hidden requirements*.

As a next step, the causes within these categories could be analyzed. Again, the diagnostic inference could be used for selecting a category and looking at the causes that usually happen within it leading to the problem being analyzed. This step is shown in Fig. 4 for the *people* category. It is possible to observe that the highest probabilities are given for causes "missing qualification of RE team members" (59%), "missing domain knowledge" (20%), and "lack of experience of RE team members" (10%).

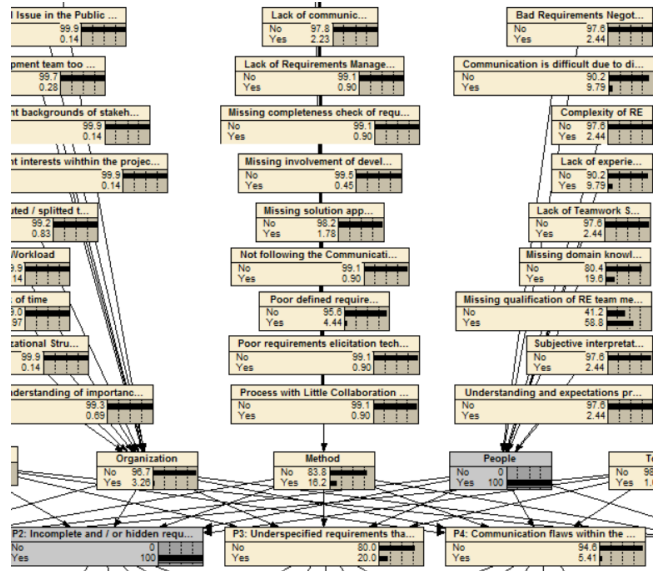

Figure 4. Diagnostic inference for causes in the *people* category leading to problem *incomplete/hidden requirements*.

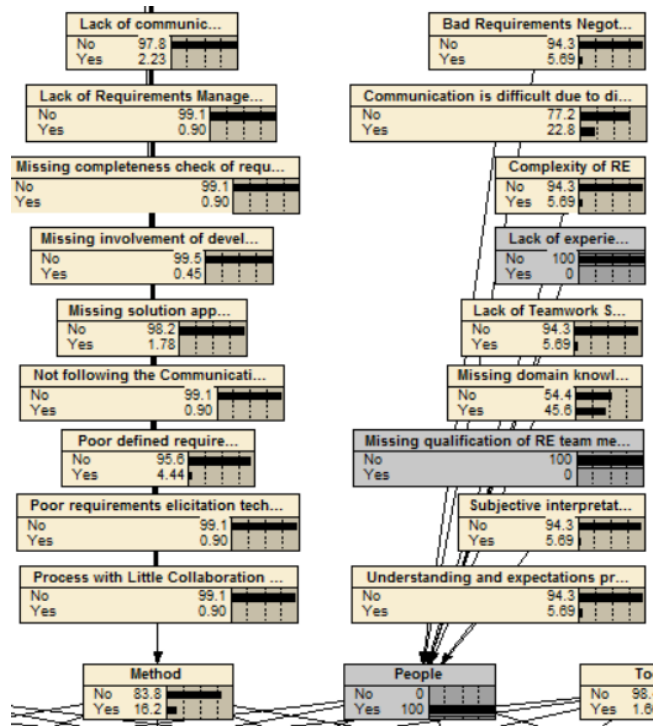

Figure 5. Diagnostic inference with additional expert knowledge indicating that two of the causes did not occur.

These and other causes could then be discussed and the Bayesian network's inference could be tailored based on this discussion. For instance, if the session participants know that the RE team members are highly qualified and experienced this could be informed to the network and the inference probabilities would be recalculated considering this knowledge. The update for this scenario is shown in Fig. 5. It can be seen that considering this expert knowledge the highest probabilities are now related to "missing domain knowledge" (46%) and "communication is difficult due to different languages" (23%).

It is noteworthy that the discussion should not be limited to one category, but focus on the identification of the main causes, which may come from different categories. For instance, for that same example, the second most likely category was *input*, in which the cause "missing engagement from customer side" appeared with the highest diagnostic probability (42%).

At the end of each session, the Bayesian network can be updated with within-company data by adding new lines at the end of the spreadsheet with the causes identified in the within-company DCA sessions and having the Bayesian network learning the probabilities again based on the new data. Thus, the cross-company data can be used as a start learning set that can be continuously improved with within-company data.

## V. INDUSTRIAL EVALUATION STRATEGY

Aiming at investigating the feasibility of using our approach in industry, we decided to adopt the empirically based technology transfer model described by Gorschek *et al.* [16]. This model is shown in Fig. 6.

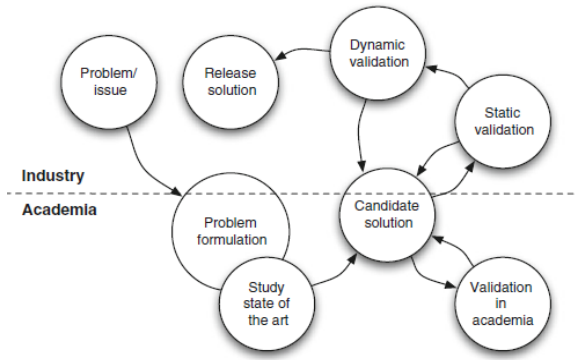

Figure 6. Technology transfer model suggested by Gorschek *et al.* [16].

Relating our research to the steps depicted in Fig. 6, the issue was supporting DCA with cross-company cause-effect data. This kind of support has not been considered so far. The candidate solution is represented by our approach (cf. Section IV). Following the model, three evaluations were conducted in different contexts: (i) in academia, (ii) with industry representatives (*static validation*), and (iii) in a real industrial case study (*dynamic validation*).

We conducted the three evaluations using the Technology Acceptance Model (TAM) [24], which has been extensively used [25]. TAM considers three constructs (*perceived usefulness*, *ease of use* and *self-predicted future use*), which are measured by a set of questions. We adapted our questions from the ones used by Ali Babar *et al.* [26]. The questions are shown in Table II. Additionally, we added questions to measure prior experience of subjects and two open text questions asking participants if they considered the cross-company data helpful for DCA and if they would change anything in our approach.

For the TAM questions we used a seven-point Likert scale (1-*completely agree, 2-largely agree, 3-partially agree, 4-neutral, 5-partially disagree, 6-largely disagree, 7-completely disagree)*, providing opportunity for optional comments on each question.

TABLE II. TAM CONSTRUCTS AND QUESTIONS

| # | Perceived Usefulness |
|---|---|
| U1 | Using the support in my job, I would be able to accomplish DCA more quickly. (Quick) |
| U2 | Using the support would improve my DCA performance. (Job Performance) |
| U3 | Using the support would increase my DCA productivity. (Increase Productivity) |
| U4 | Using the support would enhance my DCA effectiveness. (Effectiveness) |
| U5 | Using the support would make DCA easier. (Makes job easier) |
| U6 | I find the support useful to perform DCA. (Useful) |
| # | Ease of Use |
| E1 | Learning to operate the support would be easy for me. (Easy to learn) |
| E2 | I would find it easy to use the support to do what I want it to do. (Controllable) |
| E3 | My interaction with the support would be clear and understandable. (Clear and understandable) |
| E4 | It would be easy to become skillful in using the support. (Skillful) |
| E5 | It would be easy to remember how to perform DCA tasks using the support. (Remember) |
| E6 | I find the support easy to use. (Easy to use) |
| # | Self-predicted future use |
| S1 | Assuming the support is available on my job, I predict that I will use it on a regular basis in the future. |
| S2 | I prefer using the support for conducting DCA than not using it. |

## VI. INITIAL EVALUATION IN ACADEMIA

The academic evaluation took place in the context of a graduate course on software engineering during the second semester of 2015 at the University of Salvador (Brazil).

The study had 15 participants. For simulating the DCA sessions, we divided them into three groups and prepared an environment for each group, installing Netica and opening the created Bayesian network. Before starting the session, participants had to vote and agree on three of the five RE problems represented in the Bayesian network. Thereafter they were asked to discuss the main causes, based on their prior experience, using the support of the cross-company model and its diagnostic inferences during 30 minutes.

The experience of the participants in industry, with RE and the number of software projects they participated in are shown in Table III. Only one of the subjects did not have prior experience in industry and only two had no prior experience with RE in particular.

After conducting the cross-company model supported DCA session, they were asked to answer the TAM questionnaire. It is noteworthy that we could not assess the results of the DCA session itself, given that they were based on the participants' experience. Our focus was on the acceptance of the technology to support the discussions. The results of the TAM questionnaire are shown in Table IV. This table shows the frequency count for each of the seven-point Likert scale judgements, highlighting the mode (most frequently chosen judgement). It can be seen that the overall results are mainly positive for all three TAM constructs, *usefulness, ease of use* and *self-predicted use*.

TABLE III. EXPERIENCE OF THE ACADEMY PARTICIPANTS.

| # | Group | Exp. in Industry (years) | Exp. with RE (years) | # Projects |
|---|---|---|---|---|
| 1 | A | 2 | 1 | 1 |
| 2 | A | 0 | 0 | 0 |
| 3 | A | 10 | 0 | 4 |
| 4 | A | 3 | 1 | 2 |
| 5 | A | 6 | 6 | 8 |
| 6 | B | 4 | 1 | 1 |
| 7 | B | 15 | 12 | 15 |
| 8 | B | 5 | 5 | 5 |
| 9 | B | 3 | 3 | 7 |
| 10 | B | 3 | 3 | 4 |
| 11 | C | 7 | 6 | 10 |
| 12 | C | 8 | 8 | 6 |
| 13 | C | 7 | 7 | 10 |
| 14 | C | 5 | 5 | 10 |
| 15 | C | 3 | 0.5 | 1 |

TABLE IV. TAM QUESTION LIKERT SCALE FREQUENCY COUNT.

|  | 1 | 2 | 3 | 4 | 5 | 6 | 7 | Total |
|---|---|---|---|---|---|---|---|---|
| U1 | 7 | 6 | 2 | 0 | 0 | 0 | 0 | 15 |
| U2 | 7 | 2 | 5 | 1 | 0 | 0 | 0 | 15 |
| U3 | 7 | 3 | 3 | 2 | 0 | 0 | 0 | 15 |
| U4 | 8 | 4 | 3 | 0 | 0 | 0 | 0 | 15 |
| U5 | 11 | 4 | 0 | 0 | 0 | 0 | 0 | 15 |
| U6 | 12 | 1 | 2 | 0 | 0 | 0 | 0 | 15 |
| E1 | 3 | 6 | 3 | 1 | 2 | 0 | 0 | 15 |
| E2 | 4 | 9 | 2 | 0 | 0 | 0 | 0 | 15 |
| E3 | 6 | 5 | 3 | 1 | 0 | 0 | 0 | 15 |
| E4 | 8 | 5 | 1 | 1 | 0 | 0 | 0 | 15 |
| E5 | 9 | 5 | 1 | 0 | 0 | 0 | 0 | 15 |
| E6 | 9 | 5 | 1 | 0 | 0 | 0 | 0 | 15 |
| S1 | 9 | 4 | 2 | 0 | 0 | 0 | 0 | 15 |
| S2 | 4 | 6 | 2 | 1 | 1 | 1 | 0 | 15 |

Although being minority, isolated negative evaluations were observed for questions E1 and S2. In question E1, concerning learning to operate the support, the grades 5 provided by participants #2 and #9 came with textual justifications that allowed us understanding that there was a misinterpretation. Participant #2 mentioned "I would probably not be able to build the Bayesian network" and participant #9 mentioned "This is the first time I am using Netica". However, our question focused on operating the support, not on building it.

For question S2, concerning the preference of using the support, participant #7 and #12 rated it with grades 5 and 6, respectively. Their answers to the open text question on if they would change anything on the approach provided hints to interpret their judgement. Participant #7 argued "Although the provided knowledge is excellent in terms of content, I believe that the Bayesian network could be more clearly presented". Participant #12 mentioned "I would include filtering mechanisms allowing to isolate specific causes of the problem and their effects". Thus, while they would use the approach (both answered 1 to question S1), they had suggestions on the improvement of the Bayesian network presentation to practitioners.

The usefulness was also highlighted by the participants in the open text question on if they considered the cross-company data helpful. All the participants agreed in their answers, except of one who left the question empty. The answer of the most experienced participant (#7) reflects the general observed opinion of the participants well "The provided data is very important, it reflects a large range of situations, this facilitates the identification of common causes faced on a day by day basis during the professional activities. DCA can benefit from a solid knowledge base, acquired based on real situations and real problems".

VII. EVALUATION WITH INDUSTRY REPRESENTATIVES

The evaluation with industry representatives was conducted with three professionals of the Fraunhofer Project Center (FPC) at UFBA enrolled in a large-scale project, called RESCUER (Reliable and Smart Crowdsourcing Solution for Emergency and Crisis Management), in which high-quality RE plays a crucial role.

As suggested by Gorschek *et al.* [17], this follow-up to the academic evaluation was an off-line static validation with the industry representatives, i.e., we did not use the approach within the real project lifecycle to analyze causes of real defect data at this point. The main difference from the academic evaluation was that the industry representatives were active professionals that had a common project context to refer to while applying the approach.

Before starting the session, participants had to decide on three of the RE problems represented in the Bayesian network that they considered relevant for their context. Thereafter, they were asked to discuss the main causes based on their current project experience using support of the cross-company model during 30 minutes.

The experience of the participants with RE, the number of projects and their experience at the FPC are shown in Table V. It can be seen that all three are highly experienced with RE. The short time at the FPC reflects that this center was recently established in Salvador.

TABLE V. EXPERIENCE OF THE FPC INDUSTRY REPRESENTATIVES.

| # | Exp. with RE (years) | # Projects | Exp. at FPC (years) |
|---|---|---|---|
| 1 | 8 | 12 | 1 |
| 2 | 15 | 10 | 1 |
| 3 | 10 | 12 | 0.25 |

After conducting the cross-company model supported DCA session, they were asked to answer the TAM questionnaire. As this was an informal DCA session, conducted without real project data, our focus was not on the results (identified causes), but on evaluating the acceptance of using the approach, now from the perspective of industry representatives enrolled in a real project. The results of the TAM questionnaire are shown in Table VI.

The highlighted mode frequencies indicate that the industry representatives had a slightly more skeptical view than the students on the acceptance of the technology. However, overall the evaluation mainly indicates usefulness, ease of use and self-predicted future use. The only negative answer came from participant #1 on question S2. When asked on if he would change anything on the approach, he mentioned "identifying the causes is not our main problem, but finding solutions to address them". In fact, the identification of the solution option actions is currently out of the scope of the cross-company model.

TABLE VI. TAM QUESTION LIKERT SCALE FREQUENCY COUNT.

|    | 1 | 2 | 3 | 4 | 5 | 6 | 7 | Total |
|----|---|---|---|---|---|---|---|-------|
| U1 | 2 | 1 | 0 | 0 | 0 | 0 | 0 | 3 |
| U2 | 1 | 2 | 0 | 0 | 0 | 0 | 0 | 3 |
| U3 | 1 | 0 | 1 | 1 | 0 | 0 | 0 | 3 |
| U4 | 1 | 1 | 1 | 0 | 0 | 0 | 0 | 3 |
| U5 | 1 | 2 | 0 | 0 | 0 | 0 | 0 | 3 |
| U6 | 1 | 2 | 0 | 0 | 0 | 0 | 0 | 3 |
| E1 | 1 | 2 | 0 | 0 | 0 | 0 | 0 | 3 |
| E2 | 1 | 0 | 2 | 0 | 0 | 0 | 0 | 3 |
| E3 | 1 | 1 | 1 | 0 | 0 | 0 | 0 | 3 |
| E4 | 1 | 1 | 1 | 0 | 0 | 0 | 0 | 3 |
| E5 | 1 | 1 | 1 | 0 | 0 | 0 | 0 | 3 |
| E6 | 1 | 0 | 2 | 0 | 0 | 0 | 0 | 3 |
| S1 |   | 2 | 1 | 0 | 0 | 0 | 0 | 3 |
| S2 | 1 | 0 | 1 | 0 | 1 | 0 | 0 | 3 |

Concerning the open question on if they considered the cross-company data helpful for DCA all three participants agreed. Participant #1, for instance, mentioned "I agree. Mainly because if we wouldn't have the suggested causes we could possibly limit our analysis to the causes that we remember, not considering relevant causes that may have happened in our context. The data supports the analysis, making it quicker and more effective".

## VIII. INDUSTRIAL CASE STUDY

The final evaluation, as suggested by the technology transfer model [17], was an on-line dynamic validation in an industrial case study applying our cross-company data supported DCA approach within a real project lifecycle to analyze causes of problems related to real defects. The case study description is based on available guidelines [27].

### A. Case Study Objective

The case study investigates the acceptance of using our cross-company data supported DCA approach for conducting DCA sessions within a real software development project. Besides the acceptance, this time we were also interested in the results of the DCA session (e.g., identified causes).

### B. Project Context

The opportunity of conducting the evaluation in a real software development project took place at the Brazilian National Development Bank (BNDES). BNDES is the main financing agent for development in Brazil, financing important investments for the country's growth.

The goal of the project, called Garantias (*Assurances*), is to develop software to manage information on the assurances offered by organizations to BNDES as a counterpart to financial support. Assurances refer to debt acquired when financial support is approved and aim at providing security to the fulfillment of the contractual obligations. Thus, the project concerns part of BNDES's core business.

The project was divided in two modules, each one developed using an incremental and iterative lifecycle. Requirements were specified during the elaboration phase in the format of use cases. DCA was applied to defects in use case descriptions of the second module, which comprised three consecutive elaboration iterations (*EL1*, *EL2*, and *EL3*).

The requirements elicitation was conducted internally at BNDES and the specification with use cases and prototypes was conducted by employees of a third party company, working together with the internal representatives. After the specification, reviewers from BNDES performed the inspections and registered defects. The size of each of these elaboration phases in number of use cases and Function Points (FP) is shown in Table VII. This table also shows the total inspection effort and the number of defects found.

TABLE VII. SIZE OF THE ELABORATION PHASES.

| ID  | # Use Cases | Total FP | Inspection Effort (hours) | # Defects Found |
|-----|-------------|----------|---------------------------|-----------------|
| EL1 | 8           | 69       | 29                        | 69              |
| EL2 | 25          | 292      | 88                        | 181             |
| EL3 | 35          | 416      | 77                        | 214             |

The DCA team involved the BNDES Product Owner (P.O.), two BNDES reviewers, and three third party employees that worked on the use case specifications. The experience of the team members is shown in Table VIII.

TABLE VIII. EXPERIENCE OF THE DCA PARTICIPANTS.

| # | Exp. with RE (years) | # Projects | Company | Exp. at the Company (years) |
|---|----------------------|------------|---------|-----------------------------|
| 1 | 3                    | 6          | BNDES   | 8                           |
| 2 | 2                    | 2          | BNDES   | 7                           |
| 3 | 0.75                 | 1          | BNDES   | 0.75                        |
| 4 | 10                   | 10         | Third Party | 3.5                     |
| 5 | 8                    | 9          | Third Party | 2                       |
| 6 | 3                    | 3          | Third Party | 1.5                     |

### C. Instrumentation and Procedure

We were involved in the project during the *EL2* elaboration phase. As a first step, during EL2, we improved the inspection process designing a defect-based checklist (for defect categories *ambiguity, extraneous information, inconsistent information, incorrect fact,* and *omission*) and a spreadsheet to register the defects and effort according to the categories, and introducing inspection validation meetings. As we wanted to understand the whole picture, we retrospectively classified the defects found during the *EL1* phase and also registered them into the spreadsheet. For iterations *EL2 and EL3* the spreadsheet was directly used during the inspections. At the end of EL3 we had all the defect data and found the opportunity of conducting DCA.

To prepare for the DCA session, we accomplished the first three steps of our approach, *selecting the sample of the*

*defects*, *classifying selected defects* and reading their descriptions for *identifying systematic errors*. Due to space constraints, we describe how these steps were conducted for the largest iteration, *EL3*.

When *selecting the sample of defects*, we analyzed each sample using two U-charts (statistical process control charts suggested by the guidelines [5] that show varying control limits, considering the size of the sample, to help understanding defect metrics, given that defects follow a Poisson distribution), one concerning the number of defects per FP and the other with the number of defects per inspection hour. The charts allowed identifying use cases with higher defect rates and comparing the quality of the requirements of the different elaboration phases. The U-chart with the number of defects per FP for each of EL3's 35 use cases is shown in Fig. 7. It can be seen that the average number of defects per FP was 0.5 and that some use cases concentrated more defects. Later, discussions would clarify that those were the use cases with the most complex business rules.

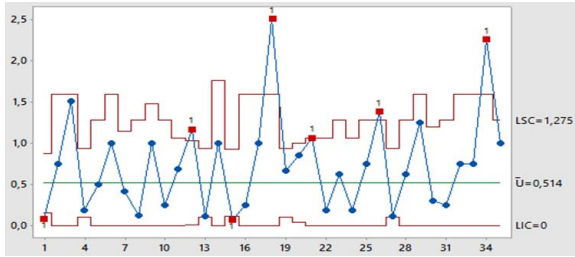

Figure 7. Defects per FP per use case of iteration *EL3*.

As the total number of defects was manageable, we used all the defects as sample for each iteration and plotted Pareto charts on the defect categories. Concerning, *classifying the selected defects*, the classification into the categories was already happening during the inspections in which the defects were revealed. The Pareto chart for *EL3* can be seen in Fig. 8. In this iteration, the defects of categories *omission* and *incorrect fact* (which together represented more than 60% of the defects) were selected for *identifying systematic errors*.

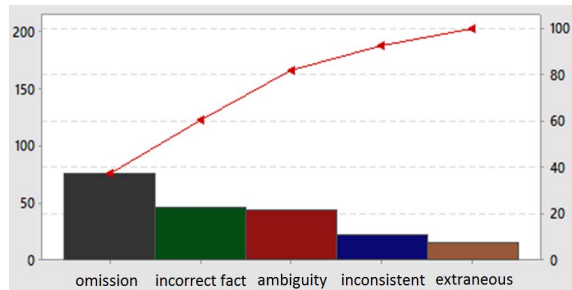

Figure 8. Pareto chart on the defects of iteration *EL3*.

To identify the systematic errors we read through the 76 defects of type omission and 46 incorrect facts trying to understand what exactly has been omitted or misunderstood. Grouping those defects into clusters helped identifying the systematic errors for which the causes should be determined in the DCA session. The listing of all types of omissions and incorrect facts that happened more than 5 times is shown in Table IX. The main systematic errors found for each iteration and the number of defects produced by the error are shown in Table X.

TABLE IX. OMISSIONS AND INCORRECT FACTS IN DETAIL.

| Type | Detail | Count |
|---|---|---|
| Omission | Business rules | 11 |
| Omission | Link to business rules | 10 |
| Omission | Actor | 10 |
| Omission | Details in the prototype | 10 |
| Omission | Field of a form | 7 |
| Omission | Identification of mandatory fields | 5 |
| Incorrect Fact | Wrong understanding (comm. problem) | 19 |
| Incorrect Fact | Linking the wrong business rule | 6 |
| Incorrect Fact | Wrong the wrong use case flow | 6 |
| Incorrect Fact | Prototype is wrong | 5 |

TABLE X. MAIN SYSTEMATIC ERRORS.

| Iteration | Defect Category | Systematic Error (number of defects) |
|---|---|---|
| EL1 | Ambiguity | Underspecifying Reqs. (7) |
| EL1 | Omission | Omitting links to between use cases (5) |
| EL2 | Omission | Omitting links to Business Rules (21) |
| EL2 | Omission | Omitting details of Business Rules (7) |
| EL2 | Inc. Fact | Linking Business Rules incorrectly (7) |
| EL3 | Omission | Omitting details of Business Rules (11) |
| EL3 | Omission | Omitting links to Business Rules (10) |
| EL3 | Inc. Fact | Incorrect facts due to comm. prob. (19) |
| EL3 | Inc. Fact | Linking Business Rules incorrectly (6) |

For *determining the main causes*, the systematic errors were taken to the DCA meeting together with the samples of defects related to them. In the meeting, they were discussed between the first two authors and the six DCA participants. Those discussions were supported by the defect descriptions and the diagnostic inferences of the Bayesian cross-company cause-effect model. During this process, we referred to diagnostic inferences for common causes of three RE problems: *incomplete/hidden requirements*, *underspecified requirements* and *communication problems*.

After determining the main causes of the systematic errors, the final steps of the approach were accomplished by *developing action proposals* and *documenting the meeting results*.

### D. Results

The participants were confident in having determined the main causes for the systematic errors. The main causes are shown in Table XI. The causes included in the cross-company cause-effect model are highlighted in bold.

Some causes appearing for systematic errors of all iterations could be associated to the fact that this was the first DCA meeting. Some causes, however, were informally addressed over time (mainly based on the inherent learning involved when conducting inspections), resulting in improvements throughout the iterations. In fact, the defect rates reduced from 1 defect per FP in *EL1* to 0.6 defects per FP in *EL2* and 0.5 defects per FP in *EL3* (reduction of 50%). The efficiency of the inspections remained almost stable, with a slight improvement from finding 2.4 defects per hour in EL1, 2.1 in EL2 and 2.8 in *EL3*.

It is noteworthy that the skills of the RE team members seem to have improved during the iterations, resulting in a reduction on ambiguities (which were the main defect category only in *EL1*). Also, while peer reviews were already used in *EL1*, in EL2 and EL3 inspection meetings were introduced. Finally, during the discussions, it was mentioned that *EL2* and *EL3* had inherently more complex use cases when compared to *EL1*, which explains why incorrect fact was a relevant category for those iterations. It was also mentioned that during *EL3* changes of requirements were more frequent.

TABLE XI. MAIN CAUSES OF THE SYSTEMATIC ERRORS.

| Systematic Error | Main Causes |
|---|---|
| Underspecifying Requirements | • Absence if inspection (validation) meetings during EL1<br>• **Lack of experience of RE team members**<br>• **Lack of time** |
| Omitting links to between use cases | • Absence if inspection (validation) meetings during EL1<br>• **Lack of experience of RE team members**<br>• **Oversight** |
| Omitting links to Business Rules | • Business rules created for the first module by other team members<br>• Huge amount of Business Rules<br>• **Lack of domain knowledge** |
| Omitting details of Business Rules | • **Complexity of the domain**<br>• **Lack of domain knowledge**<br>• **Missing completeness check** |
| Linking Business Rules incorrectly | • Business rules created for the first module by other team members<br>• Huge amount of Business Rules<br>• **Oversight** |
| Incorrect facts due to comm. problems | • Changes in the team between iterations<br>• Changes in requirements in *EL3*<br>• **Complexity of the domain**<br>• **Lack of domain knowledge** |

The action proposals suggested for addressing the causes were: (i) having the P.O. providing additional training on the business domain, (ii) providing an overview presentation on the already documented business rules to make RE team members aware of them, (iii) publishing the most frequent systematic errors and examples of related defects, and (iv) improving the inspection checklist to assure that defects associated to systematic errors are captured.

Since *EL3* was the last iteration of the project, the effectiveness of these actions cannot be directly measured. However, as mentioned by one of the participants (#2) "the identified causes and action proposals will help us to take the improvements achieved throughout the iterations of the Garantias project to an organizational level".

The answers of the participants to the TAM questions is shown in Table XII. This table shows an overall scenario of acceptance regarding usefulness, ease of use and self-predicted future use. The only slightly negative answer was provided by participant #6 concerning question S2, indicating that he would probably not prefer using the support, contrasting the opinion of the other participants. Unfortunately, no further explanations were provided.

As in the other two studies, in the open question concerning if the cross-company data was helpful all participants agreed, despite of one who did not answer the question. Participant #2 mentioned "Yes, I believe that the causes are common and the statistics help a lot in the analysis", referring to the diagnostic probabilities of the Bayesian network. Participant #4 also elaborated on the answer "Yes, the comparison during the discussions facilitates identifying the causes". We are currently refining the DCA approach description to make it available to them together with the Bayesian network support.

TABLE XII. TAM QUESTION LIKERT SCALE FREQUENCY COUNT.

|    | 1 | 2 | 3 | 4 | 5 | 6 | 7 | Total |
|----|---|---|---|---|---|---|---|-------|
| U1 | 1 | 4 | 1 | 0 | 0 | 0 | 0 | 6 |
| U2 | 2 | 4 | 0 | 0 | 0 | 0 | 0 | 6 |
| U3 | 1 | 4 | 1 | 0 | 0 | 0 | 0 | 6 |
| U4 | 2 | 1 | 3 | 0 | 0 | 0 | 0 | 6 |
| U5 | 1 | 3 | 2 | 0 | 0 | 0 | 0 | 6 |
| U6 | 3 | 3 | 0 | 0 | 0 | 0 | 0 | 6 |
| E1 | 4 | 1 | 1 | 0 | 0 | 0 | 0 | 6 |
| E2 | 1 | 5 | 0 | 0 | 0 | 0 | 0 | 6 |
| E3 | 4 | 2 | 0 | 0 | 0 | 0 | 0 | 6 |
| E4 | 2 | 4 | 0 | 0 | 0 | 0 | 0 | 6 |
| E5 | 3 | 3 | 0 | 0 | 0 | 0 | 0 | 6 |
| E6 | 3 | 3 | 0 | 0 | 0 | 0 | 0 | 6 |
| S1 | 2 | 3 | 1 | 0 | 0 | 0 | 0 | 6 |
| S2 | 2 | 2 | 1 | 0 | 1 | 0 | 0 | 6 |

## IX. THREATS TO VALIDITY

Concerning the threat categories described by Runeson et al. [27], *external validity* was enhanced by using carefully analyzed NaPiRE data for building the cross-company model and by following the model for technology transfer [16], which involved conducting studies in academy, with industry representatives and in a real industrial case study. It is noteworthy that the cross-company data was judged helpful by all participants.

Concerning *construct validity,* we evaluated acceptance by using the TAM model, which has been widely used for this purpose [25]. Moreover, we triangulated the perception of the TAM questionnaire with open questions for further and comparative qualitative analyses.

To mitigate threats to *internal validity,* we conducted the studies with direct supervision of the researchers to allow observing any potential confounding factors. Finally, *reliability* of the last study may be slightly compromised given that two researchers participated in the DCA session. However, this was needed since it was a real industry partner and participants were not experienced in conducting DCA. We encourage other researchers to replicate the study with their industry partners.

## X. CONCLUDING REMARKS

This research represents the first investigation on using cross-company data to support DCA. To this end, we defined an approach that uses diagnostic inferences of a

Bayesian network, built based on cross-company data, during DCA.

To evaluate our approach, we applied the model for technology transfer suggested by Gorschek *et al.* [16]. We conducted three consecutive evaluations in different contexts: (i) in academia, (ii) with industry representatives enrolled in a real software development project at the Fraunhofer Project Center at UFBA, and (iii) in an industrial case study at the Brazilian National Development Bank were real defect data was collected and analyzed.

The results on the acceptance of the technology were very positive and the cross-company data was unanimously considered helpful for determining main causes by the participants of all three evaluations. Therefore, the main takeaways of this research are: (i) the results indicating that cross-company data can be useful to support DCA sessions, and (ii) the cross-company data supported DCA approach.

Future work comprises further extending the approach and its cross-company model (for instance, with data from organizations of other countries and context factors), and conducting replications of the industrial case study with other industry partners.

ACKNOWLEDGMENT

We would like to thank BNDES, the Fraunhofer Project Center at UFBA and the NaPiRE team. Thanks also for the support of the Brazilian Research Council (CNPq projects #460627/2014-7 and #458261/2014-9).